\documentclass[11pt]{article}
\usepackage[utf8]{inputenc}
\usepackage{amssymb,amsmath}
\usepackage{bm} 
\usepackage{booktabs} 
\usepackage{array}
\usepackage{latexsym}
\usepackage{graphicx}
\usepackage{color}
\usepackage{datetime}
\usepackage[nosort]{cite}
\usepackage{verbatim}
\usepackage{enumerate}
\usepackage{chngpage} 
\usepackage{mathrsfs}
\usepackage{euscript}
\usepackage{psfrag}

\usepackage{datetime}

\usepackage[nosort]{cite}
\usepackage{chngpage} 
\usepackage{setspace}
\usepackage{tensor}
\usepackage{physics}
\usepackage{tensor}
\newcommand{\ind}[1]{\indices{#1}}

\usepackage{mciteplus}

\usepackage[colorlinks=true,      linkcolor=blue,      urlcolor=blue,      
filecolor=blue,      citecolor=blue,       pdfstartview=FitH,     
pdfpagemode=UseNone,      bookmarksopen=true]{hyperref}  
\usepackage[all]{hypcap}     

\definecolor{cardinal}{rgb}{0.6,0,0}
\definecolor{darkgreen}{rgb}{0,0.4,0}
\definecolor{golden}{rgb}{0.92, 0.7, 0}
\definecolor{midnight}{rgb}{0, 0, 0.5}
\definecolor{darkblue}{rgb}{0, 0, 0.7}
\definecolor{purple}{rgb}{0.5, 0, 0.5}

\def\Neql#1{{\cal N}\!=\!{#1}}
\def\coeff#1#2{\relax{\textstyle \frac{#1}{#2}}\displaystyle}

\def\IR{\mathbb{R}}

\def\IT{\mathbb{T}}

\def\cD{{\cal D}}

\def\cL{{\cal L}}
\def\cM{{\cal M}}

\def\cX{{\cal X}}

\def\nBPS#1{$\frac{1}{#1}$-BPS}

\numberwithin{equation}{section}

\begin{document}

\phantom{AAA}
\vspace{-10mm}

\vspace{1.9cm}

\begin{center}

{\huge {\bf   The Special Locus }}\\

{\huge {\bf \vspace*{.25cm}  }}

\vspace{1cm}

{\large{\bf { Tobi Ramella $^{1}$  and  Nicholas P. Warner$^{1,2,3}$}}}

\vspace{1cm}

\centerline{$^1$Department of Physics and Astronomy}
\centerline{and $^2$Department of Mathematics,}
\centerline{University of Southern California,} 
\centerline{Los Angeles, CA 90089, USA}

\vspace{1cm}

$^3$Institut de Physique Th\'eorique, \\
Universit\'e Paris Saclay, CEA, CNRS,\\
Orme des Merisiers, Gif sur Yvette, 91191 CEDEX, France \\[12pt]

\vspace{10mm} 
{\footnotesize\upshape\ttfamily  ramella  @ usc.edu\,, \quad  warner @ usc.edu} \\

\vspace{1.5cm}
 
\textsc{Abstract}

\end{center}

\noindent
The {\it special locus} plays an important role in the construction of the non-BPS microstate geometries known as microstrata.  These supergravity solutions are dual to combinations of left-moving and right-moving momentum states in the D1-D5 CFT and because supersymmetry is broken the anomalous dimensions of these states are not protected.  This means even the simplest combinations of excitations can create a cascade of frequency dependences through the non-linearities of the supergravity interactions.  Solutions on the special locus manage to lock some of these anomalous dimensions together and allow one to construct complete   solutions using gauged supergravity in three dimensions.   In the dual holographic CFT, the special locus has been shown to correspond to creating a ``pure'' gas of single particle states, however, in supergravity the special locus remains mysterious especially because it does not seem to be defined by a geometric symmetry.  In this paper we reveal the supergravity structure of the special locus, first in three-dimensional supergravity and then in the uplift to six dimensions and IIB supergravity.  The key insight is that, in three dimensions, a family of dual vector fields must vanish, and this implies that  there are algebraic relations  between tensor gauge fields in six and ten dimensions. These insights show how one can generalize the special locus Ansatz to more general mode excitations of six-dimensional supergravity.  We also construct the full six-dimensional uplift of the simplest special locus.

\begin{adjustwidth}{3mm}{3mm} 
 
\vspace{-1.2mm}
\noindent

\end{adjustwidth}

\thispagestyle{empty}
\newpage


\tableofcontents

\section{Introduction}
\label{sec:Intro}
 
One of the major advances in the {\it Microstate Geometry} program in the last five years has been going beyond the  supersymmetric solutions to provide  systematic constructions on non-BPS microstate geometries.   There have been two  independent approaches to doing this: One of them has been the explicit construction of non-BPS bubbled geometries  (see, for example, \cite{Bah:2021owp, Bah:2021rki, Heidmann:2021cms,Bah:2022yji,Bah:2022pdn,Heidmann:2022zyd, Bah:2023ows}), and the other has been the generalization of superstrata to non-BPS {\it microstrata}  (see \cite{Ganchev:2021pgs,Ganchev:2021ewa,Ganchev:2023sth,Houppe:2024hyj}).    The second approach, which will be the implicit focus of this paper,  leverages all the technology, and the holographic dictionary, developed for superstrata to create non-BPS states that one can track both in supergravity and in the dual CFT.   

To be more explicit, superstrata and microstrata are constructed in the gravity dual of the D1-D5 CFT, with a geometry that is asymptotic to AdS$_3$ $\times S^3$. Superstrata are \nBPS{8} states that involve families of left-moving momentum waves on the common D1-D5 direction, while the right-moving sector remains in the (supersymmetric) ground state.  Microstrata involve both left-moving and right-moving excitations of the CFT and so have no supersymmtries.   While the BPS equations for superstrata are linear \cite{Bena:2011dd, Bena:2015bea,Ceplak:2022wri}, and there is a well-developed technology for their solution, it is extremely challenging to solve the non-linear equations of motion of the six-dimensional supergravity that describe the  non-BPS microstrata.  However it was shown in \cite{Heidmann:2019xrd, Mayerson:2020tcl,Houppe:2020oqp} that a simple set of left-moving and right-moving modes on the D1-D5 system, and hence a simple family of microstrata,  could be captured {\it exactly} using a particular gauged supergravity in three dimensions.  This hugely simplifies the equations of motion, enabling the development of perturbative and numerical solutions \cite{Ganchev:2021pgs,Ganchev:2021ewa,Ganchev:2023sth,Houppe:2024hyj}.   

Despite the limitations of gauged supergravity in three dimensions, the solutions exhibit some remarkable and interesting properties.  First, and most obvious, is that they exist and are constructible.  This put to rest a serious concern that breaking supersymmetry would catastrophically destabilize classical solutions.   More significantly,  there are microstrata with long throats, which means high red-shifts, between the asymptotic region and where they cap-off smoothly.  Thus the microstate geometry program can indeed be extended to non-supersymmetric microstructure.  Microstrata also exhibit a spectrum whose frequencies depend on the amplitudes and dynamics of the excitations. This is an essential indicator of how the spectrum of microstrata will become chaotic. Indeed, the frequency shifting was analyzed further in  \cite{Houppe:2020oqp}, where it was also suggested that a generic microstratum  would almost certainly become time-dependent and chaotic as a result of the non-linear interactions of the excitations as one deforms away from supersymmetry. 

A surprising  aspect of the analysis of microstrata in \cite{Heidmann:2019xrd, Mayerson:2020tcl,Houppe:2020oqp} was the discovery of the ``{\it Special Locus.}''
The Ans\"atze employed in \cite{Ganchev:2021pgs,Ganchev:2021ewa,Ganchev:2023sth} have time-dependent scalar fields but, in a variation of the Q-ball trick,  they required time-{\it independence} in the metric and electromagnetic currents.  To achieve this outcome, one must lock the non-trivial {\it moding} of certain  scalar fields together in a two-to-one ratio.   The amplitudes of these  phase-locked fields are, {\it a priori}, completely independent.   However,  requiring smoothness in the perturbative expansion of the non-BPS solution requires that the {\it amplitudes} of these independent fields  are also locked to one another.  It is these constraints on amplitudes that define the  {\it Special Locus}.  Away from the special locus, singularities appear at higher orders in perturbative solution and this seemed to be a consequence of imposing the time-independence in the metric and currents.    This was substantiated in \cite{Houppe:2020oqp}, where it was shown that away from the special locus, the scalar fields all develop  independent anomalous dimensions, and this ``detunes'' any attempt at phase-locking.  With this detuning,  the non-linear interactions in supergravity create a cascade of frequency dependence in all the fields in the solution, pushing the physical solution outside the Q-ball/coiffuring Ansatz.  Thus, away from the special locus, imposing the phase-locking inherent in the Q-ball/coiffuring Ansatz results in a singular solution.  

These  results led to a number of  questions.  What is the physics of the special locus?  Does it have generalizations that go beyond the simple examples studied in \cite{Ganchev:2021ewa,Ganchev:2023sth}?  More broadly, can one generalize the Ans\"atze used in \cite{Ganchev:2021pgs,Ganchev:2021ewa,Ganchev:2023sth,Houppe:2024hyj} to six-dimensional supergravity and study more general microstrata that go well beyond the limitations of gauged, three-dimensional supergravity?  The purpose of this paper is to give some, at least partial, answers to these questions.  

The special locus has been given a very nice interpretation from the CFT perspective in \cite{Ganchev:2023sth}, where it was shown that the constraint between the amplitudes results in a multi-particle CFT state that is constructed {\it purely} from a gas of single-particle constituents.  Moving off the special locus corresponds to allowing  mixing of this state with states created by CFT operators with different trace structures, and it is these operators that develop different anomalous dimensions away from the special locus.  

From the supergravity perspective, the special locus remains something of a mystery. The problem is that the solution has very little symmetry, and the specialness of the locus does not appear to be  related to any obvious physical symmetry. The solution was constructed in three-dimensional gauged supergravity, and it was discovered empirically \cite{Ganchev:2023sth} that the special locus involves some simplifications in equations of motion of the three-dimensional gauge fields: many of the vector fields satisfy  first order differential equations rather than the more complicated second order equations that one would expect from Einstein-Maxwell-Chern-Simons theory.   Our goal here is to give a more universal understanding of what these first order equations mean from the perspective of both the three-dimensional Einstein-Maxwell-Chern-Simons theory  and the six-dimensional supergravity.  
In three dimensions we characterize  the special locus in terms of vanishing dual gauge fields in the Chern-Simons theory.

The relationship of the three-dimensional and six-dimensional formulations was mapped out in \cite{Mayerson:2020tcl}.  However, because of the lack of symmetry,   the details of the six-dimensional solution,  and especially the six-dimensional metric, are extremely complicated.   The six-dimensional metric has only been explicitly computed in a supersymmetric limit in \cite{Ganchev:2022exf}, where it was shown to involve an ambi-polar, hyper-K\"ahler  metric that is related to an elliptically deformed supertube.  Nevertheless, here we are able to show how the vanishing of the three-dimensional fields that characterizes the special locus uplifts to a very simple identity between tensor gauge fields in the six-dimensional supergravity, and hence in IIB supergravity.  We also calculate the complete uplift of all the tensor gauge fields for a the simplified special locus that was the focus of \cite{Ganchev:2021pgs,Ganchev:2021ewa}.   

A by-product of the work presented here is that it involves some very non-trivial tests of the uplift formulae of \cite{Mayerson:2020tcl}, and confirms their validity as far as microstrata are concerned,

The core message of this paper is  that the special locus, which plays such an important role in microstrata, may not have much symmetry as we usually understand it, but  it does have a  remarkable simplicity in its gauge sector.    We start in Section \ref{sec:3dSugr} by reviewing the relevant version of three-dimensional, gauged supergravity. In Section \ref{sec:speciallocus}  we review the special locus, as it was defined in \cite{Ganchev:2021pgs,Ganchev:2021ewa,Ganchev:2023sth,Houppe:2024hyj}, and show how most of the first order equations that characterize the special locus can be captured by the vanishing of a gauge connection in three dimensions.  We summarize the relevant aspects of the six-dimensional uplift in Section \ref{sec:6Duplift}, and show how the special locus is also characterized by an identity between six-dimensional fluxes. In Section \ref{sec:simplest} we obtain a complete construction of all the six-dimensional fluxes on a simplified special locus, as we make some final remarks in Section  \ref{sec:Conclusions}.

\def\muscal{{\rho}}
\def\param1{{\sigma}} 
\def\scrC{{ \mathscr{C}}}
\def\RAdS{{R_{AdS}}}
\def\RAdSsq{{R^2_{AdS}}}

\section{The three-dimensional gauged supergravity}
\label{sec:3dSugr}

Our discussion largely follows that of \cite{Ganchev:2022exf, Ganchev:2023sth}.  Further details may be found in  \cite{Houppe:2020oqp,Ganchev:2021pgs}. 
		
\subsection{The field theory content}
\label{sec:fields}
		
The theory of interest is $\Neql{4}$ gauged supergravity theory in three dimensions.  The spectrum consists of a  graviton, four gravitini, $\psi_\mu^A$, two sets of six vector fields, $A_\mu^{IJ}= -A_\mu^{JI}$ and $B_\mu^{IJ}= -B_\mu^{JI}$, $20$ fermions, $\chi^{\dot A r}$, and  $20$ scalars parametrized by the coset: 
\begin{equation}
\frac{G}{H} ~\equiv~ \frac{SO(4,5)}{SO(4) \times SO(5)} \,.
\label{coset1}
\end{equation}
The gauge group is an $SO(4) \ltimes \IT^6$ subgroup of $G$ and the $\IT^6$ gauge invariance  is typically fixed by setting six of the scalars to zero.   The  associated gauge fields, $B_\mu^{IJ}$, can then be integrated out.  The result is an action for the graviton, the gravitini, the fermions, the  $SO(4)$ gauge fields, $A_\mu^{IJ}$, and $14$ scalars.  Here we will indeed fix the $\IT^6$ gauge invariance but we will retain the gauge fields, $B_\mu^{IJ}$, as they will play an important role in characterizing the special locus.

Capital Latin indices, $I,\,J$, denote the vector representation of $SO(4)$ and we use small Greek letters, $\mu,\,\nu$, for spacetime indices. After the $\IT^6$ gauge fixing, the remaining  scalars can be described by the non-compact generators of a  $GL(4,\IR)$ matrix, ${P_I}^J$, and an $SO(4)$ vector,  $\chi_I$.  The  gauge symmetry $SO(4) \subset G$  acts on the $\chi_I$ and on the left of  ${P_I}^J$, along with a composite, local symmetry $SO(4) \subset H$ that acts on the fermions and on the right of ${P_I}^J$. The composite local symmetry is typically fixed by eliminating the compact generators of  $P$, thereby rendering it symmetric.  Indeed, once all this gauge fixing is done, the scalars are usually parametrized by a manifestly symmetric matrix, $m_{IJ}$, and its inverse, $m^{IJ}$: 
\begin{equation}
m_{IJ}   ~\equiv~   \big(P  \, P^T\big)_{IJ} \,, \qquad m^{IJ}   ~=~  \big ( (P^{-1})^T\,P^{-1}  \big)^{IJ}  \,.
\label{Mdefn}
\end{equation}

The gauge covariant derivatives are defined in terms of the $SO(4)$-dual vectors:
\begin{equation}
{\widetilde A_\mu}{}^{IJ}  ~\equiv~ \coeff{1}{2} \,\epsilon_{IJKL}\,{A_\mu}^{KL} \,,  
\label{dualGFs}
\end{equation}
with minimal couplings 
\begin{equation}
\cD_\mu \, \cX_{I}  ~=~ \partial_\mu \, \cX_{I}   ~-~  2\, g_0\,\widetilde A_\mu{}^{IJ} \, \cX_{J} \,. 
\label{covderiv}
\end{equation}
In particular, one has:
\begin{equation}
\cD_\mu m_{IJ}   ~=~   \partial_\mu m_{IJ}  ~-~   2\, g_0\,\widetilde A_\mu{}^{IK} m_{KJ }~-~   2\, g_0\,\widetilde A_\mu{}^{JK} m_{IK }   \,   \,.
\label{Dmform}
\end{equation}

The field strengths are thus
\begin{equation}
F_{\mu \nu}{}^{IJ}  ~=~ \coeff{1}{2}\, \epsilon_{IJKL} \,  \widetilde F_{\mu \nu}{}^{KL}    ~=~  \partial_{\mu}   A_{\nu}{}^{IJ}   ~-~  \partial_{\nu}   A_{\mu}{}^{IJ}   ~-~ 2 \,  g_0 \, \big(   A_{\mu} {}^{IL} \,\widetilde  A_{\nu} {}^{LJ}  ~-~ A_{\mu} {}^{JL} \,\widetilde  A_{\nu} {}^{LI}\big)  \,.
\label{fieldstrength}
\end{equation}
The gauge coupling has dimensions of inverse length, and is related to the charges of the D1-D5 system via: 
\begin{equation}
g_0 ~\equiv~ (Q_1 Q_5)^{-\frac{1}{4}} \,.
\label{g0reln}
\end{equation}

It is convenient to define:
\begin{equation}
Y_{\mu \, IJ}  ~\equiv~   \chi_J \,\cD_\mu \chi_I ~-~  \chi_I \,\cD_\mu\chi_J \,.
  \label{Ydefinition}
\end{equation}
%

\subsection{The bosonic action}
\label{sec:3Daction}

Following \cite{Houppe:2020oqp},  we will use a metric signature\footnote{Section 2.7 of \cite{Houppe:2020oqp} discusses how to convert to the mostly positive signature.} of $(+- -) $. The action can be written as  \cite{Mayerson:2020tcl,Houppe:2020oqp}\footnote{Note that the first reference uses different metric conventions.}: 
\begin{equation}
\begin{aligned}
\cL ~=~ & -\coeff{1}{4} \,e\,R    ~+~ \coeff{1}{8}\,e \, g^{\mu \nu} \, m^{IJ}   \, (\cD_\mu\, \chi_{I})  \, (\cD_\nu\, \chi_{J})  ~+~  \coeff{1}{16}\,e \, g^{\mu \nu} \,  \big( m^{IK} \, \cD_\mu\, m_{KJ}  \big)   \big( m^{JL} \, \cD_\nu\, m_{LI}  \big)  \\
&-~   \coeff{1}{8}\, e \, g^{\mu \rho}  \, g^{\nu \sigma} \, m_{IK} \,m_{JL}\,  F_{\mu \nu }^{IJ}  \, F_{\rho \sigma }^{KL}  ~-~e\, V   \\
&+~  \coeff{1}{2}\,e  \, \varepsilon^{\mu \nu \rho} \, \Big[  g_0 \,\big(A_\mu{}^{IJ}\, \partial_\nu  \widetilde A_\rho{}^{IJ}  ~+~\coeff{4}{3}\,  g_0 \, A_\mu{}^{IJ} \,  A_\nu{}^{JK}\, A_\rho{}^{KI} \,\big) ~+~  \coeff{1}{8}\,  {Y_\mu}{}^{IJ}  \, F_{\nu \rho}^{IJ} \Big] \,,
\end{aligned}
\label{eq:3Daction}
\end{equation}
where the scalar potential is given by:
\begin{equation}
V ~=~  \coeff{1}{4}\, g_0^2   \,  \det\big(m^{IJ}\big) \, \Big [\, 2 \,\big(1- \coeff{1}{4} \,  (\chi_I \chi_I)\big)^2    ~+~ m_{IJ} m_{IJ}  ~+~\coeff{1}{2} \,  m_{IJ} \chi_I \chi_J  ~-~\coeff{1}{2} \,  m_{II}  \,  m_{JJ}\, \Big] \,.
\label{potential1}
\end{equation}
Our  conventions for $\varepsilon$ can be found in Appendix A of \cite{Ganchev:2021iwy}.

It is frequently convenient to  fix the local $SO(4)$ gauge symmetry by diagonalizing $P$ in terms of four scalar fields, $\muscal_i$:
\begin{equation}
P  ~=~  {\rm diag} \big(\,  e^{\muscal_1} \,, \,  e^{\muscal_2} \,, \,  e^{\muscal_3} \,, \,  e^{\muscal_4} \, \big) \,.
\label{Pdiag}
\end{equation}
The potential then reduces to:
\begin{equation}
\begin{aligned}
V~=~ & \coeff{1}{4}\, g_0^2   \,e^{-2\, (\muscal_1 +\muscal_2+\muscal_3+\muscal_4)} \, \Big [\, 2 \,\big(1- \coeff{1}{4} \,  (\chi_I \chi_I)\big)^2    ~+~ \big( e^{4\, \muscal_1}+e^{4\, \muscal_2}+e^{4\, \muscal_3}+e^{4\, \muscal_4}  \big)  \\
& \qquad\qquad\qquad\qquad \qquad\qquad~+~\coeff{1}{2}\, \big( e^{2\, \muscal_1}\, \chi_1^2 +e^{2\, \muscal_2}\, \chi_3^2 + e^{2\, \muscal_3}\, \chi_3^2+e^{2\, \muscal_4}\, \chi_4^2  \big) \\
& \qquad\qquad\qquad\qquad \qquad\qquad~-~\coeff{1}{2}\, \big( e^{2\, \muscal_1} +e^{2\, \muscal_2} + e^{2\, \muscal_3} + e^{2\, \muscal_4}  \big)^2  \, \Big] 
\,.
\end{aligned}
\label{potential2}
\end{equation}

The supersymmetry leads to a superpotential:
\begin{equation}
\begin{aligned}
W  ~\equiv~ & \coeff{1}{4} \, g_0 \,  (\det(P))^{-1}  \,   \Big [\, 2 \,\Big(1- \coeff{1}{4} \,  (\chi_A \chi_A)\Big) ~-~ {\rm Tr}\big(P\, P^T\big)  \, \Big] \\
~=~ & \coeff{1}{4} \, g_0 \,e^{-\muscal_1 -\muscal_2-\muscal_3-\muscal_4} \, \Big [\, 2 \,\Big(1- \coeff{1}{4} \,  (\chi_A \chi_A)\Big) ~-~ \Big( e^{2\, \muscal_1}+e^{2\, \muscal_2}+e^{2\, \muscal_3}+e^{2\, \muscal_4}  \Big) \, \Big]   \,,
\end{aligned}
\label{superpot}
\end{equation}
and  the potential may be written as
\begin{equation}
V~=~ \delta^{ij} \frac{\partial W}{\partial \muscal_i}  \frac{\partial W}{\partial \muscal_j}  ~+~ 2\, m^{IJ} \, \frac{\partial W}{\partial \chi_I} \frac{\partial W}{\partial \chi_J}   ~-~2\, W^2  \,.
\label{potential3}
\end{equation}

The potential has a supersymmetric critical point at  $\muscal_j = \chi_I =0$,  where $V$ takes the value
\begin{equation}
V_0 ~=~    -  \coeff{1}{2}\, g_0^2   \,.
\label{susypt}
\end{equation}
Setting all the other fields to zero, the Einstein equations give: 
\begin{equation}
R_{\mu \nu} ~=~ - 4 \, V_0 \, g_{\mu \nu} ~=~    2 \, g_0^2 \, g_{\mu \nu}  \,,
\label{susyvac}
\end{equation}
and the supersymmetric vacuum\footnote{One should note that because we are using a metric signature $(+ - - )$ the cosmological constant of AdS is positive, contrary to the more standard and rational choice of signature.} is an AdS$_3$ of radius, $g_0^{-1}$.  We therefore define
\begin{equation}
\RAdS ~=~ \frac{1}{g_0} \,,
\label{RAdSscale}
\end{equation}
and we will henceforth use this to set the overall scale of the metric.

\subsection{The metric}
\label{sec:metric}

The coordinate conventions are inherited from the study of superstrata in six dimensions. In particular, $(u,v)$ 
are double null coordinates that are related to the time and circle coordinates, $(t,y)$, via:
\begin{equation}
u ~\equiv~\frac{1}{\sqrt{2}} \, \big( t ~-~y  \big) \,, \qquad  v ~\equiv~\frac{1}{\sqrt{2}} \, \big(t ~+~y )\,, 
\label{uvtyreln}
\end{equation}
where $y$ is periodically identified as 
\begin{equation}
y ~\equiv~ y ~+~ 2 \pi \,R_y\,.
\label{yperiod}
\end{equation}
It is  convenient to compactify the radial coordinate, $\rho$, of AdS$_3$ and use  the scale-free coordinates:
\begin{equation}
\xi ~=~\frac{\rho}{\sqrt{\rho^2+1}} \,,  \qquad  \tau~=~ \frac{t}{R_y}\,  \,,  \qquad  \psi~=~ \frac{\sqrt{2}\, v }{R_y}\equiv\tau+\sigma\,, 
\label{xidef}
\end{equation}
where $ 0 \le \xi < 1$, $\psi$ inherits the periodicity $\psi \equiv \psi + 2 \pi$ from (\ref{yperiod}), and we have introduced the coordinate $ \sigma~\equiv~ \sigma + 2\,\pi$.

As noted in \cite{Houppe:2020oqp,Ganchev:2021pgs}, the most general three-dimensional metric can then be recast in the form:
\begin{equation}
ds_{3}^{2}  ~=~  \RAdSsq \, \bigg[ \,\Omega_1^{2} \, \bigg(d \tau +   \frac{k}{(1- \xi^{2})} \, d\psi \bigg)^2~-~\,\frac{\Omega_0^{2}}{(1-\xi^{2} )^{2}} \, \big( d \xi^2 ~+~ \xi^2 \, d \psi^2 \big) \, \bigg] \,,
\label{genmet1}
\end{equation}
for three arbitrary functions $\Omega_0$,  $\Omega_1$ and $k$ of the three coordinates, $(\tau ,\xi,\psi)$.   

Part of metric regularity is to require that there are no closed time-like curves.  In particular, this means that the $\psi$-circle must always be space-like, and hence 
\begin{equation}
\xi^2\, \Omega_0^{2}  ~-~  \Omega_1^{2}  \,k^2 ~\ge~  0 \,.
\label{CTC-bound}
\end{equation}

If one returns to the coordinates $(t,\rho,v)$, one obtains the more canonical superstratum metric:
\begin{equation}
ds_{3}^{2}  ~=~   \RAdSsq \, \bigg[ \, \frac{\Omega_1^{2}}{R_y^2} \, \Big(dt  + \sqrt{2} \, (\rho^2  + 1 ) \, k  \, dv  \Big)^2~-~  \Omega_0^{2}\,\bigg(\frac{d\rho^2}{\rho^2 + 1} ~+~\frac{2}{R_y^2} \,\rho^2\,(\rho^2 + 1) \, dv^2 \bigg)  \, \bigg]\,.
\label{genmet2}
\end{equation}
If one further sets:
\begin{equation}
\Omega_0 ~=~ \Omega_1 ~=~ 1\,, \qquad  k ~=~ \xi^2    ~=~  \frac{  \rho^2 }{ (\rho^2+ 1) }  \,,
\label{AdSvals}
\end{equation}
then (\ref{genmet2}) becomes the metric of global AdS$_3$:
\begin{equation}
ds_{3}^{2}  ~=~ \RAdSsq \, \bigg[ \,   \big(\rho^2 +1\big)\,  d\tau^2~-~  \frac{d\rho^2}{\rho^2 + 1}  ~-~ \rho^2\,  d\sigma^2 \, \bigg] \,.
\label{AdSmet}
\end{equation}
%

\subsection{The generalized microstratum truncation}
\label{sub:qball}

The Ansatz for a simple microstratum  is very easily stated.  One  requires   all the fields to be invariant under the following symmetries:
\begin{itemize}
\item[(i)] Translation  invariance under:  $\tau\to \tau -\alpha $ and   $\psi \to \psi - \beta$
\item[(ii)] Reflection invariance under $\psi \to - \psi $, $\tau \to - \tau $, accompanied by a discrete internal $SO(4)$ rotation,  $2 \to -2$, $4 \to -4$.
\end{itemize}

It may, at first sight, seem that this Ansatz is  devoid of any wave-like excitations. This is not true: the microstratum oscillations are concealed in choices of gauge.  The scalar fields are required to vanish at infinity, but the $\tau$ and $\psi$ components of the gauge fields, $\tilde A_\mu^{IJ}$, are allowed to go to  constant values at infinity. These constants can be gauged to zero by introducing $\tau$ and $\psi$ oscillations in the scalar fields.   The resulting scalar modes are thus determined by the charges of the scalars.  This is the ``phase locking'' of scalar modes in the ``Q-ball'' \cite{Coleman:1985ki} and ``coiffuring'' Ans\"atze \cite{Bena:2013ora,Bena:2014rea,Bena:2015bea,Bena:2017xbt}.  In either gauge,    the currents and energy-momentum tensor are independent of $\psi$ and $\tau$.     For the $\psi$ and $\tau$ invariant gauge choice that we are going to use, the wave modes, and hence any supersymmetry breaking, are determined by the boundary values of the $\tilde A^{IJ}$ and of the scalar fields.    This is discussed extensively in \cite{Ganchev:2023sth}.

Whatever the gauge choice, regularity of the modes at the origin requires that the constants in $\tilde A_\psi^{IJ}$, or the $\psi$ mode numbers, be quantized.   The constants in $\tilde A_\tau^{IJ}$, or the $\tau$ mode numbers, represent energies of the states.  Together these modes, or constant terms in  $\tilde A_\mu^{IJ}$, represent both left-moving and right-moving waves in an asymptotically-AdS microstratum.    

We now impose symmetries (i) and (ii) on the three-dimensional fields.     One should  note that since we are reducing the fields to the singlet sector of a symmetry action, the resulting truncation is  necessarily consistent with the equations of motion.   

First, the invariance under $\tau$- and $\psi$-translations  means that $\Omega_0$,  $\Omega_1$  and  $k$, can only depend on $\xi$.  For the fields, $\chi_I$, these symmetries imply
\begin{equation}
\chi_1 ~=~    \chi_1 (\xi)  \,, \quad \chi_3 ~=~    \chi_3 (\xi) \,, \qquad  \chi_2 ~=~   \chi_4 ~=~ 0 \,.   \label{trunc1}
\end{equation}
The scalar matrix, $m$, must take the form:
\begin{equation}
m ~=~\begin{pmatrix}
m_1 &0  & m_5 & 0\\
0 & m_2 & 0 & m_6\\
m_5 & 0 & m_3 & 0\\
0 & m_6 & 0 & m_4
\end{pmatrix} \,,
\label{mmatrix}
\end{equation}
and the $m_j$ can only be  functions of $\xi$.

The gauge fields, $\tilde A^{IJ}$, are classified as to whether they are even or odd under $2 \to -2$, $4 \to -4$:  if they are odd, they can only have components along $d\tau$ or $d\psi$, and if they are even then they can only have $d\xi$ components.  We therefore have:
\begin{equation}
\begin{aligned}
\tilde A^{12} ~=~& \frac{1}{g_0} \,\big[\,  \Phi_1(\xi)  \, d\tau ~+~  \Psi_1(\xi)  \, d\psi \, \big]\,, \qquad  \tilde A^{34} ~=~ \frac{1}{g_0} \,\big[\,\Phi_2(\xi)  \, d\tau ~+~  \Psi_2(\xi)  \, d\psi    \, \big] \,,\\
\tilde A^{23} ~=~& \frac{1}{g_0} \,\big[\,  \Phi_3(\xi)  \, d\tau ~+~  \Psi_3(\xi)  \, d\psi \, \big]\,, \qquad  \tilde A^{14} ~=~ \frac{1}{g_0} \,\big[\,\Phi_4(\xi)  \, d\tau ~+~  \Psi_4(\xi)  \, d\psi    \, \big] \,, \\
\tilde A^{13} ~=~& \frac{1}{g_0}  \, \Psi_5(\xi)  \, d\xi \,, \qquad \qquad\qquad\qquad\quad \tilde A^{24}  ~=~ \frac{1}{g_0}  \, \Psi_6(\xi)  \, d\xi   \,.
\end{aligned}
\label{gauge_ansatz}
\end{equation}
Note that we have  introduced explicit factors of $g_0^{-1}$ so as to cancel the $g_0$'s in the minimal coupling and thus render the fields and interactions scale independent. 

There are residual gauge invariances that allow one to make $\xi$-dependent $U(1)$ rotations in both the $(1,3)$ and $(2,4)$ directions. There are two natural ways to fix the gauge:

\medskip
\leftline{\underline{Diagonal gauge}} 
\medskip
Here one fixes the gauge by setting $m_5 = m_6 =0$ in (\ref{mmatrix}), and one can parametrize the matrix, $P$, by taking
\begin{equation}
P  ~=~  {\rm diag} \big(\,  e^{\frac{1}{2}(\mu_1 + \lambda_1)} \,, \,  e^{\frac{1}{2}(\mu_1 - \lambda_1)} \,, \,  e^{\frac{1}{2}(\mu_2 + \lambda_2)}   \,, \, e^{\frac{1}{2}(\mu_2 - \lambda_2)} \big)  \quad {\rm with } \quad m    ~=~   P  \, P^T \,,
\label{Pdiag-Dgauge}
\end{equation}
as in (\ref{Mdefn}).  The gauge with diagonal $P$ is more convenient for the analysis of the supersymmetry. In this gauge, the Ansatz  involves  the following nineteen arbitrary functions of  one variable, $\xi$:
\begin{equation}
{\cal F} ~\equiv~ \big\{\, \chi_1  \ \,, \chi_3  \,,\  \mu_1   \,, \  \mu_2  \,, \ \lambda_1   \,, \ \lambda_2 \,,  \  \Phi_1\,,  \dots, \Phi_4 \,, \ \Psi_1\,, \dots\,,   \Psi_6   \,, \  \Omega_0  \,, \  \Omega_1    \,, \ k \, \big\} \,.
\label{functionlist1}
\end{equation}

\newpage
\bigskip
\leftline{\underline{Axial gauge}} 
\medskip
Here one fixes the residual gauge invariance by setting $\Psi_5 = \Psi_6 =0$ in (\ref{gauge_ansatz}).  It is then useful to parametrize the matrix, $m$, in this gauge according to:
\begin{equation}
m ~=~\begin{pmatrix}
e^{\mu_1 + \lambda_1}  &0  & m_5 & 0\\
0 & e^{\mu_1 - \lambda_1} & 0 & m_6\\
m_5 & 0 & e^{\mu_2 + \lambda_2} & 0\\
0 & m_6 & 0 & e^{\mu_2 - \lambda_2} 
\end{pmatrix} \,.
\label{mmatrix-axgauge}
\end{equation}
The Ansatz now  involves  the following nineteen arbitrary functions of  one variable, $\xi$:
\begin{equation}
{\cal F} ~\equiv~ \big\{\, \chi_1  \ \,, \chi_3  \,,\  \mu_1   \,, \  \mu_2  \,, \ \lambda_1   \,, \ \lambda_2  \,, \ m_ 5  \,, \ m_6 \,,  \  \Phi_1\,,  \dots, \Phi_4 \,, \ \Psi_1\,, \dots\,,   \Psi_4   \,, \  \Omega_0  \,, \  \Omega_1    \,, \ k \, \big\} \,.
\label{functionlist2}
\end{equation}

In this paper we will follow  \cite{Ganchev:2023sth} and work in the axial gauge.

\section{The ``special locus''}
\label{sec:speciallocus}

In supergravity, the special locus is defined by a very specific algebraic relationship between the scalars $\chi_I$, the scalars $m_{IJ}$ and the Maxwell fields, $F_{\mu \nu}{}^{IJ}$ \cite{Ganchev:2023sth}.  

First, $\chi_I$  is an eigenvector of $m_{IJ}$, and the corresponding eigenvalue, $m_0$, is the only non-trivial eigenvalue of $m_{IJ}$, with 
\begin{equation}
	m_0=1-\frac{1}{2}\,\chi_I \chi_I \,.
	\label{eigenvalue}
\end{equation}
All other eigenvalues of $m_{IJ}$ are equal to $1$.    Indeed,  $m_{IJ}$, has the form
\begin{equation}
m ~=~\begin{pmatrix}
1-\frac{1}{2}\,\chi_1^2&0  & -\frac{1}{2}\,\chi_1 \chi_3 & 0\\
0 & 1 & 0 & 0\\
 -\frac{1}{2}\,\chi_1 \chi_3 & 0 & 1-\frac{1}{2}\,\chi_3^2  & 0\\
0 & 0 & 0 & 1
\end{pmatrix} \,,
\label{mmatrix2}
\end{equation}

This means that the special locus involves thirteen functions.
\begin{equation}
{\cal F} ~\equiv~ \big\{\, \chi_1  \ \,, \chi_3  \,,  \  \Phi_1\,,  \dots, \Phi_4 \,, \ \Psi_1\,, \dots\,,   \Psi_4   \,, \  \Omega_0  \,, \  \Omega_1    \,, \ k \, \big\} \,.
\label{functionlist3}
\end{equation}

On the special locus,  $\chi_I$ is a null vector of the dual field strengths.  
\begin{equation}
	\widetilde F_{\mu \nu}{}^{IJ} \,  \chi_J ~=~ 0 \,.
\label{nullvec}
\end{equation}

These two conditions seem to suggest an $SO(3)$ invariance in the directions orthogonal to $\chi_I$, but it is not that simple.  The  Maxwell fields and their gauge  connections, $\tilde A_{\mu}^{IJ}$, are  non-trivial in the directions orthogonal to $\chi_I$, and therefore are not invariant under this $SO(3)$.  In six-dimensions these gauge connections define the non-trivial fibration of an $S^3$ over the three-dimensional metric, and the six-dimensional metric is extremely non-trivial.  It seems we must look elsewhere for simplicity. 
 
\subsection{The algebraic and first-order equations on the special locus}
\label{ss:SLeqns}

The condition (\ref{nullvec}) imposes the following algebraic constraints on the Maxwell fields:
\begin{equation}
\Phi_3\,\Psi_1 ~+~ \Phi_4\,\Psi_2~-~\Phi_1\,\Psi_3~-~\Phi_2\,\Psi_4 ~=~0 \,,
\label{nullvec2a}
\end{equation}
as well as requiring:
\begin{equation}
\chi_1\,\Phi_1' ~-~ \chi_3\,\Phi_3'  ~=~ 0\,,\qquad \chi_3\,\Phi_2' ~+~ \chi_1\,\Phi_4'~=~ 0 \,, \qquad
\chi_1\,\Psi_1' ~-~ \chi_3\,\Psi_3'~=~ 0\,,\qquad \chi_3\,\Psi_2' ~+~ \chi_1\,\Psi_4'~=~0 \,.
\label{nullvec2b}
\end{equation}

To write down the other equations of motion, it is useful to define
\begin{equation}
\Lambda=2\,m_0~\equiv~ 2~-~ \big(\chi_1^2+\chi_3^2\big)\,, \quad H ~\equiv~  \frac{\xi^2 \, \Omega_0^2}{(1-\xi^2)^2 \Omega_1^2}    \,, \quad F_j   ~\equiv~ \Psi_j - \frac{k}{1- \xi^2}\,  \Phi_j \,,  \quad  j =1,\dots,4  \,.
\label{FHdefns}
\end{equation}

If one substitutes these into the equations of motion one then finds that the latter are satisfied by the following first-order equations:
\begin{equation}
\begin{aligned}
\xi \partial_\xi \Phi_1 &~=~ - \frac{\Omega_1 \, \chi_3}{\Lambda} \,\big( \chi_1\, F_4  ~+~\chi_3\, F_2\big)\,, \qquad   \xi \partial_\xi \Phi_2 ~=~ - \frac{\Omega_1 \, \chi_1}{\Lambda} \,\big( \chi_1\, F_1  ~-~\chi_3\, F_3\big)  \\
\xi \partial_\xi \Phi_3& ~=~ - \frac{\Omega_1 \, \chi_1}{\Lambda} \,\big( \chi_1\, F_4  ~+~\chi_3\, F_2\big) \,, \qquad \xi \partial_\xi \Phi_4  ~=~  \frac{\Omega_1 \, \chi_3}{\Lambda} \,\big( \chi_1\, F_1  ~-~\chi_3\, F_3\big)\,,
\end{aligned}
\label{Maxeqn1a}
\end{equation}
and
\begin{equation}
\begin{aligned}
\xi \partial_\xi \Psi_1 &~=~ - \frac{\Omega_1 \, \chi_3}{\Lambda} \,\bigg[ \bigg(\frac{k}{1- \xi^2}\bigg)\,\big( \chi_1\, F_4  ~+~\chi_3\, F_2\big)~+~ H\,\big( \chi_1\, \Phi_4  ~+~\chi_3\, \Phi_2\big)\bigg]\,, \\
   \xi \partial_\xi \Psi_2 &~=~ - \frac{\Omega_1 \, \chi_1}{\Lambda} \,\bigg[ \bigg(\frac{k}{1- \xi^2}\bigg)\, \big( \chi_1\, F_1  ~-~\chi_3\, F_3\big) ~+~ H\,\big( \chi_1\, \Phi_1  ~-~\chi_3\, \Phi_3\big)\bigg]\,, \\
\xi \partial_\xi \Psi_3 & ~=~ - \frac{\Omega_1 \, \chi_1}{\Lambda} \,\bigg[ \bigg(\frac{k}{1- \xi^2}\bigg)\,\big( \chi_1\, F_4  ~+~\chi_3\, F_2\big) ~+~ H\,\big( \chi_1\, \Phi_4  ~+~\chi_3\, \Phi_2\big)\bigg]\,,\\
 \xi \partial_\xi \Psi_4  &~=~  \frac{\Omega_1 \, \chi_3}{\Lambda} \,\bigg[ \bigg(\frac{k}{1- \xi^2}\bigg)\, \big( \chi_1\, F_1  ~-~\chi_3\, F_3\big) ~+~ H\,\big( \chi_1\, \Phi_1  ~-~\chi_3\, \Phi_3\big)\bigg] \,.
\end{aligned}
\label{Maxeqn2a}
\end{equation}
Note that these equations are consistent with (\ref{nullvec2b}).  

One also obtains the following equation for the scalars:
\begin{equation}
 \frac{ \Omega_1}{\Lambda} \, \xi \,\big(\chi_3\,\partial_\xi  \chi_1 - \chi_1\,\partial_\xi \chi_3\big)  ~=~ \big(\Phi_4\,\Psi_1-\Phi_1\,\Psi_4+\Phi_3\,\Psi_2-\Phi_2\,\Psi_3\big)\,,
\label{nueqn1a}
\end{equation}
as well as first-order equations for two of the metric functions:
\begin{equation}
\Omega_1 \, \xi \partial_\xi\bigg(\frac{k}{1-\xi^2}\bigg) ~+~ 8\big(\Phi_1\,\Phi_2~+~ \Phi_3\,\Phi_4~-~ \kappa_1 \big) H ~=~ 0 \,,
\label{keqn1}
\end{equation}
and
\begin{equation} 
\Omega_1 \, \xi \partial_\xi \log (H)  ~-~  8 \big(\Phi_1\,F_2+\Phi_2\,F_1+\Phi_3\,F_4+\Phi_4\,F_3\big) ~-~ 16\, \kappa_1\, \bigg(\frac{k}{1-\xi^2}\bigg) ~+~   \kappa_2 ~=~ 0\,,
\label{Heqn1}
\end{equation}
where $\kappa_1$ and $\kappa_2$ are constants of integration.

One can combine  (\ref{keqn1}) and (\ref{Heqn1}) to arrive at:
\begin{equation}
\Omega_1 \, \xi \partial_\xi \bigg(\,H ~-~ \bigg(\frac{k}{1-\xi^2}\bigg)^2 \, \bigg) ~-~  8 \big(\Phi_1\, \Psi_2+\Phi_2\,\Psi_1+\Phi_3\,\Psi_4+\Phi_4\,\Psi_3\big)\,H ~+~   \kappa_2\,H  ~=~ 0\,,
\label{Heqn2}
\end{equation}

We therefore have eleven first-order equations for the thirteen functions in (\ref{functionlist3}).   The remaining two functions can be taken to be  $\Omega_1$ and $\chi_1^2+\chi_3^2$, or $\Lambda$.  For these one needs to use the equations of motion.  These last equations can be found in \cite{Ganchev:2023sth}.   

The important point is that the complete set of equations of motion are solved if one uses the first order equations (\ref{Maxeqn1a})--(\ref{Heqn1})  combined with the two remaining equations of motion for $\Lambda$  and $\Omega_1$.   These equations were discovered empirically in \cite{Ganchev:2023sth} and are {\it sufficient} to solve the equations of motion.  There could be other solutions, but it was also shown in \cite{Ganchev:2023sth} that the smooth microstratum solution obeys these first-order equations and so it is likely that any other solutions to the equations of motion are singular. 

Our purpose here is to give some geometric understanding of at least the nine equations (\ref{Maxeqn1a})--(\ref{nueqn1a}) for the scalars and Maxwell fields.  To do this one must return to the original formulation of the three-dimensional supergravity.

\subsection{Three-dimensional supergravity revisited}
\label{ss:3Dsugr}

The important thing to recall is that, as originally formulated \cite{Cvetic:2000dm,Cvetic:2000zu, Nicolai:2001sv, Nicolai:2001ac, Nicolai:2003bp, Nicolai:2003ux,Deger:2014ofa,Samtleben:2019zrh}, the gauged supergravity actually has an extended, non-semi-simple  gauge group $SO(4) \ltimes \IT^6$ and the gauge action was written as a pure Chern-Simons action.   The action in Section \ref{sec:3dSugr}, with its Yang-Mills term, only emerges once the non-semi-simple  gauge fields, ${B_\mu}{}^{IJ}$, of  $\IT^6$ have been integrated out.   The  CS formulation made the original construction simpler in terms of generalized geometry, and, as we will see in Section \ref{ss:uplift}, the non-semi-simple  gauge fields of  $\IT^6$ appear naturally in the ``uplift formulae'' that reconstruct six-dimensional fields.

In the formulation of interest here \cite{Mayerson:2020tcl, Houppe:2020oqp}, the theory starts out with   $20$ scalar fields as well as the 
 $SO(4) \ltimes \IT^6$  gauge group.  The twelve gauge connections are denoted:
\begin{equation}
 {A_\mu}^{IJ} ~=~ - {A_\mu}^{JI}   \,, \qquad  A_{\mu}{}^{J}{}_I ~=~  -{A_{\mu \,J}}^I \,,
\end{equation}
and their duals were written:
\begin{equation}
{\widetilde A_\mu}{}^{IJ}  ~\equiv~ \coeff{1}{2} \,\epsilon_{IJKL}\,{A_\mu}^{KL} \,, \qquad \ {\widehat A_\mu}{}^{IJ}  ~\equiv~ \coeff{1}{2} \,\epsilon_{IJKL}\,{{A_\mu}{}^K}{}_L   \,.
\label{dualGFs2}
\end{equation}
It simplifies things to define: 
\begin{equation}
{B_\mu}{}^{IJ}  ~\equiv~ 4\,g_0\, \big(\, {\widetilde A_\mu}{}^{IJ}    -  {\widehat A_\mu}{}^{IJ} \big)  \,,
\label{Bvecden}
\end{equation}
and work with ${B_\mu}{}^{IJ}$ and ${\widetilde A_\mu}{}^{IJ}$ as gauge connections.    The field ${B_\mu}{}^{IJ}$ will ultimately be integrated out to create the action defined in Section \ref{sec:3dSugr}. 

After gauge fixing the $\IT^6$ by reducing the number of scalars from 20 to 14, the original three-dimensional bosonic action becomes: 
\begin{equation}
\begin{aligned}
\cL ~=~ & -\coeff{1}{4} \,e\,R   ~+~ \coeff{1}{8}\,e \, g^{\mu \nu} \, m^{IJ}   \, (\cD_\mu\, \chi_{I})  \, (\cD_\nu\, \chi_{J}) ~+~  \coeff{1}{16}\,e \, g^{\mu \nu} \,  \big( m^{IK} \, \cD_\mu\, m_{KJ}  \big)   \big( m^{JL} \, \cD_\nu\, m_{LI}  \big)   \\
 &~+~  \coeff{1}{16}\,e \, g^{\mu \nu} \,  m^{IJ} \, m^{KL} \,\big({B_\mu}{}^{IK} ~+~  \coeff{1}{2}\, Y_{\mu \, IK}\big) \,\big({B_\nu}{}^{JL} ~+~  \coeff{1}{2}\, Y_{\nu \, JL}  \big)   ~-~e\, V  \\
& ~+~  \coeff{1}{2}\,e  \, \varepsilon^{\mu \nu \rho} \, \Big[  g_0 \,\big(A_\mu{}^{AB}\, \partial_\nu  \widetilde A_\rho{}^{AB}  ~+~\coeff{4}{3}\,  g_0 \, A_\mu{}^{AB} \,  A_\nu{}^{BC}\, A_\rho{}^{CA} \,\big) ~-~  \coeff{1}{4}\,  {B_\mu}{}^{AB}  \, F_{\nu \rho}^{AB} \Big] \,.
\end{aligned}
\label{eq:3Daction2}
\end{equation}
Note that  ${B_\mu}{}^{IJ}$ appears as a massive, non-dynamical vector field and the action only has Chern-Simons interactions.

Completing the square in all the terms that involve ${B_\mu}{}^{IJ} $, one arrives at the action\footnote{There is a sign error in the Yang-Mills term in  \cite{Mayerson:2020tcl}.  This is correctly given in \cite{Houppe:2020oqp}}:
\begin{equation}
\begin{aligned}
\cL ~=~ & -\coeff{1}{4} \,e\,R  ~+~ \coeff{1}{8}\,e \, g^{\mu \nu} \, m^{IJ}   \, (\cD_\mu\, \chi_{I})  \, (\cD_\nu\, \chi_{J})  ~+~  \coeff{1}{16}\,e \, g^{\mu \nu} \,  \big( m^{IK} \, \cD_\mu\, m_{KJ}  \big)   \big( m^{JL} \, \cD_\nu\, m_{LI}  \big) \\
 & ~-~   \coeff{1}{8}\, e \, g^{\mu \rho}  \, g^{\nu \sigma} \, m_{IK} \,m_{JL}\,  F_{\mu \nu }^{IJ}  \, F_{\rho \sigma }^{KL}    ~-~e\, V ~+~ {\cal L}_{\rm B} \\
& ~+~  \coeff{1}{2}\,e  \, \varepsilon^{\mu \nu \rho} \, \Big[  g_0 \,\big(A_\mu{}^{AB}\, \partial_\nu  \widetilde A_\rho{}^{AB}  ~+~\coeff{4}{3}\,  g_0 \, A_\mu{}^{AB} \,  A_\nu{}^{BC}\, A_\rho{}^{CA} \,\big) ~+~  \coeff{1}{8}\,  {Y_\mu}{}^{AB}  \, F_{\nu \rho}^{AB} \Big] \,,
\end{aligned}
\label{eq:3Daction3}
\end{equation}
where
\begin{equation}
\begin{aligned}
{\cal L}_{\rm B} ~\equiv~ &  \coeff{1}{16}\, e \, g^{\mu \nu} \, m^{IK} \,m^{JL}\,\Big({B_\mu}{}^{IJ}  +  \coeff{1}{2}\, Y_\mu{}_{ \, IJ}   - g_{\mu \sigma_1} \, \varepsilon^{\sigma_1  \rho_1 \rho_2} \, m_{I P_1} m_{J P_2}\,F_{\rho_1 \rho_2}^{P_1 P_2}\Big) \\
 &\qquad\qquad\qquad\qquad\qquad \times  \Big({B_\nu}{}^{KL} +  \coeff{1}{2} \,Y_\nu{}_{ \, KL} - g_{\nu \sigma_2} \, \varepsilon^{\sigma_2 \rho_3 \rho_4}\, m_{K P_3} \,m_{L P_4}\,F_{\rho_3 \rho_4}^{  P_3 P_4}   \Big) \,.
\end{aligned}
\label{Baction}
\end{equation}
Integrating out the the gauge fields, ${B_\mu}{}^{IJ}$, gives their classical equations:
\begin{equation}
{B_\mu}{}^{IJ} ~=~  g_{\mu \rho} \,  \varepsilon^{\rho \sigma \nu} \, m_{IK} m_{JL}\,F_{\sigma \nu}^{KL}   ~-~ \coeff{1}{2}\,  Y_\mu{}_{ \, IJ}  \,.
 \label{Beqn}
\end{equation}
and hence $ {\cal L}_{\rm B}$ can be dropped from the action (\ref{eq:3Daction3}). The  result is the action (\ref{eq:3Daction}).  Finally, observe that using   (\ref{dualGFs2}) and  (\ref{Bvecden}),  one can recast (\ref{Beqn}) as a difference of the $\IT^6$ and $SO(4)$ connections:
\begin{equation}
g_0\, \big(\,{{A_\mu}{}^I}{}_J  ~-~ {A_\mu}^{IJ} \big)  ~=~ -\coeff{1}{8} \, g_{\mu \rho} \,  \varepsilon^{\rho \sigma \nu} \,\epsilon_{IJKL} \, m_{KP} m_{LQ}\,F_{\sigma \nu}^{PQ}   ~+~ \coeff{1}{16}\, \epsilon_{IJKL} \,  Y_\mu{}_{ \, KL}  \,.
 \label{Bdualeqn}
\end{equation}
%

\subsection{The first-order scalar-Maxwell equations of the special locus}
\label{ss:SLeqns2}

The bottom line is that the first order equations for the scalars and Maxwell fields (\ref{Maxeqn1a}),  (\ref{Maxeqn2a}) and (\ref{nueqn1a}) are completely equivalent to:
\begin{equation}
{B_\mu}{}^{IJ} ~=~ 0 \,,
 \label{Bzero1}
\end{equation}
or setting the $\IT^6$ connection equal to the $SO(4)$ connection:
\begin{equation}
{{A_\mu}{}^I}{}_J  ~-~ {A_\mu}^{IJ}  ~=~  0 \,.
 \label{Aseql1}
\end{equation}
In other words, the first-order  scalar-Maxwell equations trivialize this connection.

As we will now show, this equation has major implications for the structure of the six-dimensional solution.

\section{The six-dimensional uplift of the gauge fields}
\label{sec:6Duplift}

The uplift formulae for the metric and tensor gauge fields were obtained in  \cite{Mayerson:2020tcl}.  The details are fairly complex, but we will give a short, self-contained review here, and show how the special locus is characterized by a constant proportionality of the $D1$ and $D5$ fluxes from the six-dimensional perspective.  

We start by summarizing the technology of the uplift formula and our presentation draws heavily on  \cite{Mayerson:2020tcl} and Appendix A of \cite{Samtleben:2019zrh}, but we  fix the parameters in \cite{Mayerson:2020tcl} by taking:
\begin{equation}
\alpha ~=~ \gamma_0 ~=~     \coeff{1}{2}\, g_0  \,, \qquad \varepsilon =-1 \,,
\label{gammalphrepl}
\end{equation}
as in \cite{Houppe:2020oqp,Ganchev:2021ewa,Ganchev:2023sth}.

\subsection{The uplift formulae}
\label{ss:uplift}

\subsubsection{Some geometry on $S^3$}
\label{ss:S3geom}

The starting point is the ``round'' $S^3$, which is thought of as a unit sphere defined by Cartesian coordinates, $\mu^A$ in $\IR^4$, satisfying $\mu^A \mu^A =1$.  Following earlier work, it is convenient to coordinatize this by taking:
\begin{equation}
\mu^1 = \sin\theta \sin\varphi_1\,, \quad \mu^2 = \sin\theta \cos\varphi_1\,, \quad \mu^3 = \cos\theta \sin\varphi_2\,, \quad \mu^4 = \cos\theta\cos\varphi_2\,,
\label{CartCoords}
\end{equation}
with $y^i$ coordinates $\big(y^1, y^2, y^3\big) =(\theta, \varphi_1,\varphi_2)$
The metric on the round $S^3$  is then:
\begin{equation}
 \mathring{ds}^2_{S^3} =~  \mathring{g}_{ij}dy^idy^j ~=~ d\theta^2 ~+~  \sin^2\theta d\varphi_1^2 ~+~  \cos^2\theta d\varphi_2^2 \,.
 \label{dsRndSphere}
\end{equation}
Note that quantities labelled with $\mathring{}$ will always be those of the round sphere.  In particular, in the  $y^i$  coordinates, $(\theta, \varphi_1,\varphi_2)$, we define
\begin{equation}
\mathring{\omega}_{ijk}  ~ \equiv~ \sin\theta\cos\theta \,  \epsilon_{ijk} \,,
 \label{omodefn1}
\end{equation}
where  $\epsilon_{ijk}$ is completely antisymmetric and $\epsilon_{123}=+1$,  is the volume form on the round $S^3$.   It is convenient to define the following vector field:
\begin{equation}
\mathring{\zeta}^k ~=~ \frac{1}{2} \,  \big(\,\tan\theta\,,0\,,0\, \big)  \,.
 \label{zeta1}
\end{equation}
The important point about $\mathring{\zeta}^k$ is that $\mathring{\zeta}^k\mathring{\omega}_{ijk}$ provides a $2$-form potential for the volume form,  $\mathring{\omega}_{ijk}$:
\begin{equation}
d \Big[  \mathring{\zeta}^k  \mathring{\omega}_{ijk} \, dy^i \wedge dy^j \Big]~=~\frac{1}{2} \, d \Big[  \sin^2\theta \,  d\varphi_1 \wedge d\varphi_2 \Big] ~=~  \sin\theta \,  \cos \theta \,  d \theta \wedge d\varphi_1 \wedge d\varphi_2   \,.
 \label{2formpot}
\end{equation}

We also need the gauge covariant derivatives on the sphere :
\begin{equation}
\cD\mu^A ~\equiv~  d\mu^A~-~ 2\, g_0\,  \tilde{A}^{AB}\,\mu^B \quad \Leftrightarrow \quad \cD y^i ~\equiv~  dy^i ~-~ 2 g_0 \mathcal{K}^i_{AB} \, \tilde{A}^{AB}  \quad \Rightarrow \quad  \cD\mu^A ~=~  \partial_i\mu^A \cD y^i  \,, 
 \label{cDsphere1}
\end{equation}
where
\begin{equation}
 \mathcal{K}^i_{AB} = \mathring{g}^{ij}\partial_j \mu^{[A} \mu^{B]} \,,
\end{equation}
are the rotational Killing vectors on the sphere.  

Finally, we need the invariants: 
\begin{equation}
 \Delta ~\equiv~  m_{AB}\mu^A \mu^B\,, \qquad   X  ~\equiv~   \chi_A \, \mu^A 
 \label{eq:mDeltaDef}
 \end{equation}
and note that for $m_{IJ}$ of the form (\ref{mmatrix2}) we have:
\begin{equation}
 X  ~=~ \chi_1 \, \sin\theta \sin\varphi_1~+~  \chi_3 \,  \cos\theta \sin\varphi_2  \,,  \qquad \Delta ~=~   1 ~-~ \frac{1}{2} \, X^2 \,.  
 \label{eq:mDeltaform}
 \end{equation}
%

\subsubsection{Uplifting the scalars and metric}
\label{ss:upscalmet}

The six-dimensional metric is
\begin{equation}
ds_6^2 ~=~  (\det m_{AB})^{-1/2} \, \Delta^{1/2} \, ds_3^2  ~+~  g_0^{-2}\, (\det m_{AB})^{1/2}\, \Delta^{-1/2} \, m^{AB} \cD\mu^A \cD\mu^B,
 \label{eq:6dmet}
 \end{equation}
where $ds_3^2$ is the three-dimensional metric (\ref{genmet1}).

The six-dimensional scalars are a dilaton and axion, $\varphi$ and $X$, where the uplift of $X$ is defined above and the dilaton is given by:
\begin{equation} 
 e^{-\sqrt{2}\varphi} ~=~   \Delta \,, \qquad X  ~=~   \chi_A \, \mu^A \,.
  \label{eq:6ddil}
 \end{equation}

The six-dimensional theory has an $SO(1,2)$ covariance in the tensor gauge fields, while the scalars have a sigma model structure.  That is, the scalars fit into an $SO(1,2)$ matrix:
\begin{equation} 
 \cM\ind{^{\hat{I}}_{\hat{J}}} ~=~ \frac{1}{2}\, e^{\sqrt{2}\varphi}\,
 \left(  \begin{array}{ccc} 
 X^2 & 8 & -2\sqrt{2} X
 \\ \frac18e^{-2\sqrt{2}\varphi} (2 +  e^{\sqrt{2}\varphi} X^2)^2 & X^2 & -\frac{X}{2\sqrt{2}}(2e^{-\sqrt{2}\varphi}+X^2) \\
  +\frac{X}{\sqrt{2}}(2e^{-\sqrt{2}\varphi}+X^2) & 4\sqrt{2}X & -2e^{-\sqrt{2}\varphi} -2 X^2
  \end{array}\right), 
 \label{scalMdef}   
 \end{equation}
The $SO(1,2)$ metric, and its inverse are: 
\begin{equation}
 \eta_{{\hat{I}}{\hat{J}}} ~=~
 \left( \begin{array}{ccc} 
 0 & 1 & 0\\ 1 & 0 & 0\\ 0 & 0 & -\frac{1}{2}
 \end{array}\right)  \,,  \qquad  
 \eta^{{\hat{I}}{\hat{J}}} = 
 \left( \begin{array}{ccc} 
 0 & 1 & 0\\ 1 & 0 & 0\\ 0 & 0 & -2
 \end{array}\right)\,. 
\label{etaDefs} 
\end{equation}
Note that the scalar matrix $\mathcal{M}_{{\hat{I}}{\hat{J}}}$ (with both indices down) is symmetric.

\subsubsection{The six-dimensional equations }
\label{ss:6d-eoms}

There are three three-form field strengths $G^{\hat{I}}$, $\hat{{I}}=1,2,4\,,$\footnote{We conform to the idiosyncratic notation and conventions for the six-dimensional three-forms that is used in the superstrata literature. This slightly odd notation of omitting the index 3 is historical. In reduction of the six-dimensional system to five dimensions the gauge fields labelled by $3$, emerge as Kaluza-Klein fields.}  transforming as a vector of $SO(1,2)$.  The three-forms are all closed 
\begin{equation}
 dG^{\hat{I}}  ~=~  0 \,.
  \label{eq:6Dbianchi} 
 \end{equation}
and satisfy a self-duality relation:
\begin{equation}
\hat{*}\, G^{\hat{I}}  ~\equiv~   \frac{1}{\big(3!\big)^2} \,  \epsilon\ind{_{\hat{\mu}  \hat{\nu} \hat{\rho}}}{^{\hat{\alpha}\hat{\beta}\hat{\gamma}}} \, G^{\hat{I}}_{\hat{\alpha}\hat{\beta}\hat{\gamma}} \, dx^ {\hat{\mu}} \wedge dx^{\hat{\nu}}\wedge dx^{\hat{\rho}}  ~=~  \epsilon_6\,   \mathcal{M} \ind{^{\hat{I}}}{_{\hat{J}}} \, G^{\hat{J}} \,,
\label{eq:6Dselfduality} 
\end{equation}
where $\epsilon_6~=~ \pm 1$ and $\mathcal{M} \ind{^{\hat{I}}}{_{\hat{J}}}$ is the scalar matrix  (\ref{scalMdef}).   The combination of   (\ref{eq:6Dbianchi}) and   (\ref{eq:6Dselfduality}) are the  equations of motion of these fields.

Note that the matrix (\ref{scalMdef}) has one positive eigenvalue and two negative eigenvalues. It then follows from (\ref{eq:6Dselfduality}) that for superstrata and microstrata (with two anti-self dual tensors)  one should take $\epsilon_6=+1$, while for $\epsilon_6 = -1$, the theory has two self-dual tensors and hence the uplift formulae should reduce to a truncation of that given in \cite{Samtleben:2019zrh}.\footnote{Our conventions for the six-dimensional Hodge dual are given explicitly in (\ref{eq:6Dselfduality}). While never explicitly mentioned in \cite{Samtleben:2019zrh}, their convention for Hodge duals is such that their self-duality relation receives a relative minus sign compared to ours in (\ref{eq:6Dselfduality}). In this paper  we take the six-dimensional Levi-Civita tensor to decompose as $\epsilon_{\mu\nu\rho ijk} = + \epsilon_{\mu\nu\rho} \epsilon_{ijk}$}  Since we are interested in microstrata we take $\epsilon_6 =  + 1$.

 The other bosonic equations of motion can be obtained by varying the pseudo-Lagrangian \cite{Ferrara:1997gh,Riccioni:2001bg}:
\begin{equation} 
 \mathcal{L}_{6D}  ~=~
  R ~-~ \frac{1}{2}\, (\partial_{\hat{\mu}}\varphi)^2 ~-~ \frac{1}{2}\,  e^{\sqrt{2}\varphi}\,  (\partial_{\hat{\mu}} X)^2 ~-~ \frac{1}{6}\,  \mathcal{M}_{{\hat{I}}{\hat{J}}}\, G\ind{^{\hat{I}}_{{\hat{\mu}}{\hat{\nu}}{\hat{\rho}}}}\,  G^{{\hat{J}}{\hat{\mu}}{\hat{\nu}}{\hat{\rho}}}\,.
\label{eq:6Dlagr}
\end{equation}

\subsubsection{Uplifting the tensor gauge fields }
\label{ss:tensuplift}

The expressions for the uplift of the three-forms are partially implicit and thus rather complicated.  Specifically, one writes explicit uplift formulae for sufficient components of the two-form potentials, $B^{\hat{I}}$, so as to determine the remainder via the equations of motion (\ref{eq:6Dbianchi}) and   (\ref{eq:6Dselfduality}).

As usual, the potentials are defined as a result of  (\ref{eq:6Dbianchi}):
\begin{equation}  
G^{\hat{I}} ~=~  dB^{\hat{I}}\,.
\end{equation}

The three-forms $G^{\hat{I}}$, and their two-form potentials $B^{\hat{I}}$, can be decomposed in terms of their components along the sphere or along the three-dimensional space-time:
\begin{equation}  
\begin{aligned} 
G^{\hat{I}}  ~=~  & \ \frac{1}{3!} \, G^{{\hat{I}}}_{ijk} \, \cD y^i\wedge \cD y^j\wedge \cD y^k  ~+~  \frac{1}{2} \, G^{\hat{I}}_{ij\mu} \, \cD y^i\wedge \cD y^j\wedge dx^\mu \\
& +  \frac{1}{2} \, G^{\hat{I}}_{i\mu\nu} \,\cD y^i\wedge dx^\mu\wedge dx^\nu ~+~  \frac{1}{3!} \, G^{\hat{I}}_{\mu\nu\rho} \,dx^\mu\wedge dx^\nu\wedge dx^\rho,\\
   B^{\hat{I}} ~=~ &  \frac{1}{2} \,  B^{\hat{I}}_{ij} \cD y^i\wedge \cD y^j ~+~ B^{\hat{I}}_{i\mu} \cD y^i\wedge dx^\mu ~+~ \frac{1}{2} \, B^{\hat{I}}_{\mu\nu} dx^\mu\wedge dx^\nu .
\end{aligned}
\label{BGfields}
\end{equation}

It then suffices to  give expressions for $B^{\hat{I}}_{ij}$ and $B^{\hat{I}}_{i\mu}$, because they unambiguously determine the components $G^{\hat{I}}_{ijk},G^{\hat{I}}_{ij\mu}$ of the three-forms.  The other components $G^{\hat{I}}_{i\mu\nu},G^{\hat{I}}_{\mu\nu\rho}$ are then determined by the self-duality relation (\ref{eq:6Dselfduality}).  The remaining components, $B^{\hat{I}}_{\mu\nu}$, of the potentials can then be constructed by integration. 

The uplift formulae  for $B^{\hat{I}}_{ij}$ are:
\begin{align}
 \label{eq:6DansatzB1ij} B^1_{ij} &~=~\frac{2}{g_0^2} \,\bigg(\,\mathring{\omega}_{ijk}\mathring{\zeta}^k   ~-~  \frac{1}{4} \, \mathring{\omega}_{ijk}\,\mathring{g}^{kl}\,\Delta\,\partial_l \big[\,\Delta^{-1}\,\big]     \bigg)\,,  \\
\label{eq:6DansatzB2ij}  B^2_{ij} &~= \frac{1}{2\, g_0^2} \, \bigg(\,\mathring{\omega}_{ijk}\mathring{\zeta}^k  ~-~   \frac{1}{8} \,  \mathring{\omega}_{ijk}\,\mathring{g}^{kl}\,\Delta\,\partial_l \big[\Delta^{-1}X^2\big]    \bigg)\,, \\
\label{eq:6DansatzB4ij}   B^4_{ij} &~=~ - \frac{1}{\sqrt{2} \, g_0^2} \, \mathring{\omega}_{ijk}\,\mathring{g}^{kl}\, \Delta^{1/2}\,\partial_l \big[\,\Delta^{-1/2} X \,\big]\,.
\end{align}
while the ansatze for the components $B^{\hat{I}}_{i\mu}$ is:
\begin{align}
\label{eq:6DansatzB1imu} B^1_{i\mu} &~=~  -\frac{2}{g_0} \, \partial_i\mu^A \, A^{AB}_\mu \, \Big[ \,\mu^B -  2\, \mathring{\zeta}^k \, \partial_k\mu^B\, \Big]\,, \\
 \label{eq:6DansatzB2imu} B^2_{i\mu} &~=~ -  \frac{1}{2\, g_0} \, \partial_i\mu^A\,  \Big( \Big[ \,\tensor{A}{_{\mu}^{A}_{B}} ~-~  A_\mu^{AB}\,\Big]\mu^B ~+~ A^{AB}_\mu\Big[\,\mu^B - 2\,\mathring{\zeta}^k\,\partial_k\mu^B \, \Big]\Big)\,,  \\
  \label{eq:6DansatzB4imu} B^4_{i\mu} &= 0 \,.
\end{align}
%

\subsubsection{Implementing the uplift on the special locus }
\label{ss:impuplift}

As we remarked earlier, the non-trivial gauge fields make the metric uplift very complicated.  However, there is a remarkable simplification in the gauge sector.

First notice that  the majority of the equations that define  the special locus are embodied in (\ref{Aseql1}) and this results in a huge simplification of the expression for $B^2_{i\mu}$,  (\ref{eq:6DansatzB4imu}).

Also observe that the structure of the dilaton and axion,  (\ref{eq:mDeltaform}) give us:
\begin{equation}
\Delta\,\partial_l \big[\,\Delta^{-1}\,\big]  ~=~\Delta^{-1} X \, \partial_l X  \,, \quad\Delta\,\partial_l \big[\Delta^{-1}X^2\big]  ~=~  2\,\Delta^{-1} X \, \partial_l X\,,  \quad  \Delta^{1/2}\,\partial_l \big[\,\Delta^{-1/2} X \,\big] ~=~ \Delta^{-1} \partial_l X   \,.
\label{dilders1}
\end{equation}

We then find:
 \begin{equation}
\begin{aligned}
B^1_{ij} &~=~\frac{2}{g_0^2} \,\bigg(\,\mathring{\omega}_{ijk}\mathring{\zeta}^k   ~-~  \frac{1}{4} \, \Delta^{-1}\,  \mathring{\omega}_{ijk}\,\mathring{g}^{kl}\, X \, \partial_l X \,   \bigg)\,, \qquad  B^2_{ij}  ~= \frac{1}{4} \, B^1_{ij}\,, \\
  B^4_{ij} &~=~ - \frac{1}{\sqrt{2} \, g_0^2} \, \mathring{\omega}_{ijk}\,\mathring{g}^{kl}\,\Delta^{-1}\, \partial_l X \,.
\end{aligned}
\label{eq:6DansatzBij} 
\end{equation}
and 
\begin{equation}
B^1_{i\mu} ~=~  -\frac{2}{g_0} \, \partial_i\mu^A \, A^{AB}_\mu \, \Big[ \,\mu^B -  2\, \mathring{\zeta}^k \, \partial_k\mu^B\, \Big]\,, \qquad
B^2_{i\mu} ~=~\frac{1}{4} \, B^1_{i\mu}  \,,  \qquad 
B^4_{i\mu} = 0 \,.
\label{eq:6DansatzBimu} 
\end{equation}
We note, in passing, that 
 \begin{equation}
B^1_{ij} ~-~ \frac{X}{\sqrt{2}}\,  B^4_{ij}  ~=~\frac{2}{g_0^2} \, \mathring{\omega}_{ijk}\mathring{\zeta}^k   \,,
\label{eq:B1B4ij} 
\end{equation}
which shows that the fluctuations of $B^4$ on the sphere are intimately related to those of $B^1$.

The $SO(1,2)$ matrix becomes:
\begin{equation} 
 \cM\ind{^{\hat{I}}_{\hat{J}}} ~=~\Delta^{-1} \,
 \left(  \begin{array}{ccc} 
 \frac{1}{2}\,  X^2 & 4 & - \sqrt{2} X
 \\ \frac{1}{4} &  \frac{1}{2}\,  X^2  & -\frac{1}{2\sqrt{2}} \, X \\
\frac{1}{\sqrt{2}} \, X  & 2\sqrt{2} \, X & -(1+ \frac{1}{2}) \, X^2
  \end{array}\right), 
 \label{scalMsl}   
 \end{equation}
and that if one defines $W_{\hat{I}} = (1\,, -4\,,0)$ then:
\begin{equation} 
W_{\hat{I}}  \, \cM\ind{^{\hat{I}}_{\hat{J}}} ~=~  - W_{\hat{J}}  \,.
\label{evec1}   
 \end{equation}

It follows from (\ref{eq:6Dselfduality}) that
\begin{equation}
\hat{*}\, W_{\hat{I}} \, G^{\hat{I}}  ~=~     -  \epsilon_6\,   W_{\hat{I}} \, G^{\hat{J}}~=~     -   W_{\hat{I}} \, G^{\hat{J}}  \,,
\label{Wasdual} 
\end{equation}
where we have set $\epsilon_6 = +1$ because we use the standard superstratum conventions.

From (\ref{eq:6DansatzBij}) and  (\ref{eq:6DansatzBimu}), we know that 
$$
W_{\hat{I}} \, {G^{\hat{I}}}_{ijk} ~=~ 0 \,, \qquad W_{\hat{I}} \, {G^{\hat{I}}}_{ij\mu} ~=~ 0 \,,
$$
and hence  (\ref{eq:6Dselfduality}) tells us that all the other components of $W_{\hat{I}} \, G^{\hat{I}}$ must vanish.

We therefore conclude that on the special locus we have:
\begin{equation}
B^{{2}}  ~=~  \frac{1}{4}  \, B^{{1}}  \,, \qquad   G^{{2}}  ~=~  \frac{1}{4}  \, G^{{1}}  \,.
\label{SLB1eqlB2} 
\end{equation}

Conversely, because this identity uses all of the equations (\ref{Maxeqn1a}),  (\ref{Maxeqn2a}) and (\ref{nueqn1a}),  along with the algebraic structure of the scalar matrix $m_{IJ}$, we suspect that this identity defines that special locus.

The other left-eigenvectors of $ \cM\ind{^{\hat{I}}_{\hat{J}}}$ are $U_{\hat{I}} = (X \,,  4\, X\,,-2\sqrt{2})$ and $V_{\hat{I}} = (1 \,,  4 \,,-\sqrt{2}\,X)$ with eigenvalues $-1$ and $+1$ respectively.  These lead to the identities
\begin{equation}
\hat{*}\, \big( X \, G^{1}    ~-~ \sqrt{2}\, G^{4} \big) ~=~     -   \big( X \, G^{1}    ~-~ \sqrt{2}\, G^{4} \big) \,, \qquad \hat{*}\, \big( \sqrt{2} \, G^{1}    ~-~ X\, G^{4} \big) ~=~      \big( \sqrt{2} \, G^{1}    ~-~ X\, G^{4} \big) \,,
\label{eq:otherGs} 
\end{equation}
where we have also set  $\epsilon_6 =  + 1$

This reduces the tensor and scalar structure to a single scalar field, $X$, and one self-dual and one anti-self-dual tensor multiplet where:
\begin{equation}
\hat{*}\, \begin{pmatrix}
G^{1} \\ G^{4} 
\end{pmatrix} 
  ~=~
\frac{1}{\Delta} \begin{pmatrix}
(1 + \frac{1}{2} X^2)& -\sqrt{2} \, X \\  \sqrt{2} \, X  & -(1 + \frac{1}{2} X^2)
\end{pmatrix} \,\begin{pmatrix}
G^{1} \\ G^{4} 
\end{pmatrix}  \,,
\label{redGeqn1}
\end{equation}
The scalar matrix here belongs to $SO(1,1)$ with 
$$
\eta ~=~  \begin{pmatrix}
1& 0 \\ 0 & -1
\end{pmatrix} \,.
$$
It is evident from  (\ref{Wasdual}) that the  special locus constraint annihilates exactly one of the anti-self-dual tensor multiplets, leaving one with one self-dual and one anti-self-dual tensor multiplet. 
There is thus an $SO(1,2)$, six-dimensional duality rotation that reduces the original special locus to six-dimensional minimal supergravity coupled to a single anti-self-dual tensor multiplet.  

From the perspective of IIB supergravity,  $B^1$ and $B^2$ are the fields, $C^{(p)}$, sourced by  the $D5$ and $D1$ branes,  while $B^4$ is sourced by the  Kalb-Ramond NS field, $B^{(2)}$.  Indeed, $C^{(2)} \sim B^2$ and $C^{(6)} \sim B^1 \wedge {\rm Vol} _{\IT^4}$ and $B^{(2)} \sim B^4$.  Thus the six-dimensional duality rotation that rotates $B^1$ and $B^2$ is not a symmetry of the IIB theory.   Nevertheless, the special locus does lock the D1 and D5 sources together so that they are dual to one another modulo the volume of $\IT^4$.   Interestingly, the NS field,  $B^{(2)} \sim B^4$, is, at leading order ($X \to 0$) anti-self-dual.  This means that that $B^4$ encodes, at leading order, NS5 and F1 fluxes that are dual to one another in much the same way that the D1 and D5 fluxes are dual to one another.

\section{A simple special locus}
\label{sec:simplest}

To obtain a complete analytic uplift of the tensor gauge fields is, in principle, possible but is something of a technical challenge.  Here we will illustrate the process by  specializing to a simpler  single-family special locus that played a major role in the early work on microstrata \cite{Ganchev:2021pgs,Ganchev:2021ewa}.   The simplified special locus has an additional $U(1)$ symmetry:  rotational invariance in the $(3,4)$ directions of the gauge indices, and hence a $U(1)$ isometry on the $S^3$ of the uplift to six dimensions.

\subsection{Reducing to the simplified locus}
\label{ss:RedSimp}

\subsubsection{The equations and constraints}
\label{ss:eqnsconstr}

Imposing the $U(1)$ symmetry means that $\chi_3 \equiv 0$ and 
\begin{equation}
m ~=~\begin{pmatrix}
1-\frac{1}{2}\,\chi_1^2&0  & 0 & 0\\
0 & 1 & 0 & 0\\
 0 & 0 & 1  & 0\\
0 & 0 & 0 & 1
\end{pmatrix} \,,
\label{mmatrix3}
\end{equation}
The gauge field Ansatz, (\ref{gauge_ansatz}), reduces to:
\begin{equation}
\tilde A^{12} ~=~  \frac{1}{g_0} \,\big[\,  \Phi_1(\xi)  \, d\tau ~+~  \Psi_1(\xi)  \, d\psi \, \big]\,, \qquad  \tilde A^{34} ~=~ \frac{1}{g_0} \,\big[\,\Phi_2(\xi)  \, d\tau ~+~  \Psi_2(\xi)  \, d\psi    \, \big] \,, 
\label{gauge_ansatz2}
\end{equation}
with all other gauge fields vanishing. 

The algebraic constraint (\ref{nullvec2a}) is trivially satisfied and the first-order special locus equations reduce to:
\begin{equation}
\begin{aligned}
\xi \partial_\xi \Phi_1 &~=~ 0 \,, \qquad  \xi \partial_\xi \Psi_1 ~=~  0\,, \qquad  \Lambda~=~ 2\,m_0~\equiv~ 2~-~ \chi_1^2\,,  \\
  \xi \partial_\xi \Phi_2 & ~=~ - \frac{\Omega_1 }{\Lambda} \, \chi_1^2\, F_1    \,, \qquad  \xi \partial_\xi \Psi_2  ~=~ - \frac{\Omega_1 }{\Lambda}\, \chi_1^2 \,\bigg[ \bigg(\frac{k}{1- \xi^2}\bigg)\, F_1  ~+~ H  \, \Phi_1\bigg]     \,.
\end{aligned}
\label{Maxeqn3}
\end{equation}
with all other equations trivially satisfied.  Note that the gauge fields, $\Phi_1$ and $\Psi_1$, are constants but they are non-trivial, and, indeed, essential to creating non-trivial solutions. Also observe that (\ref{Maxeqn1a})--(\ref{nueqn1a}) still leave  non-trivial dynamics for $\Psi_2$ and $\Phi_2$  and so the flux constraint, (\ref{SLB1eqlB2}), remains  as a  non-trivial condition.

The first-order equations for the metric functions become:
\begin{equation}
\Omega_1 \, \xi \partial_\xi\bigg(\frac{k}{1-\xi^2}\bigg) ~+~ 8\big(\Phi_1\,\Phi_2 ~-~ \kappa_1 \big) H ~=~ 0 \,.
\label{keqn2}
\end{equation}
and
\begin{equation} 
\Omega_1 \, \xi \partial_\xi \log (H)  ~-~  8 \big(\Phi_1\,F_2+\Phi_2\,F_1\big) ~-~ 16\, \kappa_1\, \bigg(\frac{k}{1-\xi^2}\bigg) ~+~   \kappa_2 ~=~ 0\,,
\label{Heqn3}
\end{equation}
where $\kappa_1$ and $\kappa_2$ are constants of integration.

Again, combining  (\ref{keqn2}) and (\ref{Heqn3}) to arrive at:
\begin{equation}
\Omega_1 \, \xi \partial_\xi \bigg(\,H ~-~ \bigg(\frac{k}{1-\xi^2}\bigg)^2 \, \bigg) ~-~  8 \big(\Phi_1\, \Psi_2+\Phi_2\,\Psi_1\big)\,H ~+~   \kappa_2\,H  ~=~ 0\,.
\label{Heqn4}
\end{equation}
%

\subsubsection{Pieces of the uplift}
\label{ss:upliftbits}

First, the scalar quantities are:
\begin{equation}
  X  ~=~ \chi_1 \, \sin\theta \sin\varphi_1  \,,  \qquad \Delta ~=~   1 ~-~ \frac{1}{2} \, X^2 \,.  
 \label{eq:mDeltaform2}
 \end{equation}
Given the form of $\mathring{\zeta}^k$, (\ref{zeta1}), we also have 
$$
\mu^B - 2\,\mathring{\zeta}^k\,\partial_k\mu^B  ~=~ \frac{1}{\cos^2 \theta} \, \big( \, 0\,, 0\,, \mu^3\,, \mu^4 \,  \big)\,,
$$
and hence:
\begin{equation}
B^1_{i\mu} \, dy^i \wedge dx^\mu ~=~  -\frac{2}{g_0} \,   \partial_i\mu^A \, dy^i  \wedge A^{AB}  \, \Big[ \,\mu^B -  2\, \mathring{\zeta}^k \, \partial_k\mu^B\, \Big] ~=~  \frac{2}{g_0^2} \,   \big[\,  \Phi_1(\xi)  \, d\tau ~+~  \Psi_1(\xi)  \, d\psi \, \big] \wedge d\varphi_2
\label{eq:B1imu} 
\end{equation}
where we have used   $ A^{34} = \tilde A^{12}$ and (\ref{gauge_ansatz2}).   

Using (\ref{eq:6DansatzB1ij}),  (\ref{eq:6DansatzB4ij})  and (\ref{dilders1}), one obtains
\begin{align}
 \label{eq:B1ij}  \coeff{1}{2} \, B^1_{ij} \, dy^i \wedge dy^j  ~=~& \frac{1}{g_0^2} \,\bigg( \,\sin^2 \theta  \, d\varphi_1 \wedge d\varphi_2   ~-~ \frac{1}{2}\,  \sin \theta  \,\cos \theta \, \Delta^{-1}\,  \big( X \partial_{\theta} X\big) \, d\varphi_1 \wedge d\varphi_2  \nonumber \\
&  \qquad\qquad\qquad  \qquad   \quad \ ~+~ \frac{1}{2}  \,\cot \theta \, \Delta^{-1}\,  \big( X \partial_{\varphi_1} X\big) \, d\theta \wedge d\varphi_2    \bigg)\,,    \\
\label{eq:B4ij} \coeff{1}{2} \,  B^4_{ij} \, dy^i \wedge dy^j  ~=~ &- \frac{1}{\sqrt{2} \, g_0^2 \, \Delta} \, \bigg(\,  \sin \theta  \,\cos \theta \,  \big(\partial_{\theta} X\big) \, d\varphi_1 \wedge d\varphi_2 ~-~  \cot \theta \,\big(  \partial_{\varphi_1} X\big) \, d\theta \wedge d\varphi_2    \bigg)  \,.
\end{align}
So that
\begin{equation}
 \frac{1}{2} \, \bigg(B^1_{ij} ~-~\frac{1}{\sqrt{2}} \, X \, B^4_{ij}\bigg)\, dy^i \wedge dy^j  ~=~ \frac{1}{g_0^2} \,\sin^2 \theta  \, d\varphi_1 \wedge d\varphi_2  
\label{eq:simpIdent} 
\end{equation}

To assemble the $B$-fields, we see from (\ref{BGfields}) one must replace the differentials, $dy^i$ by their covariant forms, $\cD y^i$, which are given by (\ref{cDsphere1}):
\begin{equation}
  \cD \theta ~=~  d\theta  \,, \quad  \cD \varphi_1 ~=~  d\varphi_1 ~-~   2 \,   \big[\,  \Phi_1(\xi)  \, d\tau +  \Psi_1(\xi)  \, d\psi \, \big] \,, \quad \cD \varphi_2 ~=~  d\varphi_1 ~-~  2 \,   \big[\,  \Phi_2(\xi)  \, d\tau +  \Psi_2(\xi)  \, d\psi \, \big]
 \label{covDyi}
\end{equation}

This gives the following pieces of the tensor gauge fields:
\begin{align}
 \label{eq:B1}  B^1~=~& \frac{1}{g_0^2} \,\bigg[ \,\bigg(\sin^2 \theta    ~-~ \frac{1}{2}\,  \sin \theta  \,\cos \theta \, \Delta^{-1}\,  \big( X \partial_{\theta} X\big)\bigg) \, \cD\varphi_1 \wedge \cD\varphi_2  \nonumber \\
&  \qquad  ~+~ \frac{1}{2}  \,\cot \theta \, \Delta^{-1}\,  \big( X \partial_{\varphi_1} X\big) \, d\theta \wedge \cD\varphi_2~+~ 2 \,   \big(\,  \Phi_1(\xi)  \, d\tau  +  \Psi_1(\xi)  \, d\psi \, \big) \wedge \cD\varphi_2   \, \bigg]  ~+~ \dots \,,    \\
\label{eq:B4}  B^4 ~=~ &- \frac{1}{\sqrt{2} \, g_0^2 \, \Delta} \, \bigg[\,  \sin \theta  \,\cos \theta \,  \big(\partial_{\theta} X\big) \, \cD\varphi_1 \wedge \cD\varphi_2 ~-~ \cot \theta \,\big(  \partial_{\varphi_1} X\big) \, d\theta \wedge \cD\varphi_2   \, \bigg]~+~ \dots  \,,
\end{align}
where $+ \dots$ means components with index structure $B^{\hat{I}}_{\mu \nu}$.

It is interesting to note that, because of (\ref{Maxeqn3}), the term $2 \big(  \Phi_1(\xi) d\tau  +  \Psi_1(\xi)  d\psi  \big) \wedge  d\varphi_2$ is pure gauge.  This implies that the terms in  $2 \big(  \Phi_1(\xi) d\tau  +  \Psi_1(\xi)  d\psi  \big) \wedge  \cD \varphi_2$ that are not pure gauge have only $\mu \nu$ indices, and so one can drop this term from  (\ref{eq:B1}).  It follows that  (\ref{eq:simpIdent}) can be generalized to:
\begin{equation}
B^1~-~ \frac{X}{\sqrt{2}}\, B^4 ~=~ \frac{1}{g_0^2} \, \sin^2 \theta    \, \cD\varphi_1 \wedge \cD\varphi_2  ~+~ \dots \,,   
 \label{eq:simpIdent2} 
 \end{equation}
where, once again, $+ \dots$ means components with index structure $B^{\hat{I}}_{\mu \nu}$.  The term on the right-hand side of  (\ref{eq:simpIdent2}) is, of course, the flux that represents the pure D1 and D5 charges of the unperturbed background.  Thus, in the $S^3$ directions and mixed $S^3$ and AdS$_3$ directions, the fluctuations in $B^4$ and $B^1$ are locked together.  As we will see, this is not true of the purely AdS$_3$ components of these fields.

\subsubsection{The uplifted tensor gauge fields}
\label{ss:upliftG}

As we noted above, we can use  (\ref{eq:B1}) and  (\ref{eq:B4})  to construct the components:  $G^{\hat{I}}_{ijk}$ and $G^{\hat{I}}_{ij\mu}$. The remaining components can then be determined by duality (\ref{eq:6Dselfduality}) or, more specifically, (\ref{eq:otherGs}).  One then checks that the $G^{\hat{I}}$ are closed.   One can then construct the potentials.  This entire calculation is straightforward, but  surprisingly laborious.   The final result is deceptively simple and yields the following potentials:
\begin{align}
 \label{eq:B1fnl}  B^1~=~& 4\,B^{{2}}  ~=~ \frac{1}{2\, g_0^2 \, \Delta} \,\sin \theta \, \bigg[ \, \chi_1^2  \,\cos \theta \,  \sin \varphi_1 \,\cos \varphi_1 \,  d\theta ~+~\sin \theta \, 
 \Big(2 - \chi_1^2\, \sin^2 \varphi_1 \Big) \,     \cD\varphi_1 \,\bigg]  \wedge \cD\varphi_2    \nonumber \\
&  \qquad\qquad  \qquad~+~ K(\xi) \, d \tau \wedge d \psi \,,    \\
\label{eq:B4fnl}  B^4 ~=~ &  \frac{1}{\sqrt{2} \, g_0^2 } \, \bigg[\, \frac{\chi_1\,\cos \theta }{\Delta}\,\big( \,  \cos \varphi_1 \,  d\theta ~-~\sin \theta \, \cos \theta\, \sin \varphi_1 \, 
  \cD\varphi_1 \,\big)  \wedge \cD\varphi_2    \nonumber \\ 
 &  \qquad\qquad  ~+~ \frac{2\, \Omega_1\, \chi_1\,\sin \theta \, \cos \varphi_1}{\xi \, \big(1 - \frac{1}{2} \chi_1^2\big) } \, d\xi \wedge \bigg( F_1\,  \bigg(d \tau + \frac{k}{(1 -  \xi^2)}  \,d \psi \bigg)~+~  \frac{\xi^2 \, \Omega_0^2}{ (1 -  \xi^2)^2\, \Omega_1^2 } \, \Phi_1\, d \psi \bigg)  \nonumber \\
  &  \qquad\qquad  ~+~  \frac{\xi \, \Omega_1\,\chi_1'\,\sin \theta \, \sin \varphi_1}{\big(1 - \frac{1}{2} \chi_1^2\big) } \, d\tau \wedge d \psi \, \bigg] \,,
\end{align}
where
\begin{equation}
 K'(\xi) ~=~ -  \frac{\xi \,\Omega_0^2\, \Omega_1 \, \big( 4 - \chi_1^2\big)}{(1 -  \xi^2)^2\, \big( 2 - \chi_1^2\big)}
 \label{Fdefn1}
\end{equation}
To integrate $G^4$ to this form one needs the equation of motion for $\chi_1$:
\begin{equation}
\frac{ \xi}{\Omega_1} \,\partial_\xi \bigg( \frac{\Omega_1\, \xi \partial_\xi  \chi_1 }{\big(1 - \frac{1}{2} \chi_1^2\big)} \bigg)~+~  \frac{\Omega_1^2\, H\,  \chi_1 }{\big(1 - \frac{1}{2} \chi_1^2\big)} ~-~ 4\,\chi_1\,\bigg(  \frac{ \xi}{\Omega_1} \,  \big( \Phi_1 \partial_\xi \Psi_2  ~-~ \Psi_1 \partial_\xi \Phi_2     \big)~+~(F_1^2 ~-~H\, \Phi_1^2) \bigg) ~=~ 0\,
 \label{chi1eqn1}
\end{equation}
which, on the special locus, reduces to:
\begin{equation}
\frac{ \xi}{\Omega_1} \,\partial_\xi \bigg( \frac{\Omega_1\, \xi \partial_\xi  \chi_1 }{\big(1 - \frac{1}{2} \chi_1^2\big)} \bigg)    ~-~ \frac{4\,  \chi_1 }{\big(1 - \frac{1}{2} \chi_1^2\big)} \,\bigg(F_1^2 ~-~H \,\bigg(\Phi_1^2 + \frac{1}{4}\, \Omega_1^2\bigg) \bigg) ~=~ 0 \,.
 \label{chi1eqn2}
\end{equation}

We also note that there is a slightly simpler gauge choice that arises from using (\ref{Maxeqn3}) to establish the identity:
\begin{equation}
 \frac{2\, \Omega_1\, \chi_1\,\sin \theta \, \cos \varphi_1}{\xi \, \big(1 - \frac{1}{2} \chi_1^2\big) } \, d\xi \wedge \bigg( F_1\,  \bigg(d \tau + \frac{k}{(1 -  \xi^2)}  \,d \psi \bigg)~+~  \frac{\xi^2 \, \Omega_0^2}{ (1 -  \xi^2)^2\, \Omega_1^2 } \, \Phi_1\, d \psi \bigg)   ~=~ \frac{2\,\sin \theta \, \cos \varphi_1}{\chi_1} \, d \, \big[\, \cD\varphi_2   \, \big]
 \label{cleanergauge}
\end{equation}
The only downside of this gauge is that it may not be smooth as $\chi_1 \to 0$.

Finally, observe that the $dy^i \wedge dy^j$ and  $dy^i \wedge dx^\mu$ terms in (\ref{eq:B1fnl}) and (\ref{eq:B4fnl}) match those of (\ref{eq:B1}) and (\ref{eq:B4}), as they must. Even though $B^1$ and $B^4$ are related by a simple identity, (\ref{eq:simpIdent}), there is a significant difference in the $dx^\mu \wedge dx^\nu$ components.

We have also directly verified that $G^{\hat{I}} = d B^{\hat{I}}$ exactly reproduce (\ref{eq:otherGs}).

It is also important to note that our results here  also provide a very non-trivial test of the uplift formulae of \cite{Mayerson:2020tcl}.  These flux formulae are rather involved and, in other gauged supergravities, have been notoriously difficult to get ``right.''  It is therefore very gratifying to have such a successful implementation as the one presented here.

\section{Final comments}
\label{sec:Conclusions}

We have shown that in six-dimensional supergravity, the special locus  involves a remarkably simple self-duality constraint on the tensor gauge fields, effectively trivializing one of the anti-self-dual tensor multiplets.    This means that the D1 and D5 configurations are locked to one another by  duality.    We have also seen that, in the three-dimensional Chern-Simons gauged supergravity, the same special locus equations are characterized by the vanishing of the $\IT^6$ gauge potential or, equivalently, the vanishing of the dual Chern-Simons gauge field defined by the right-hand side of (\ref{Beqn}).  

There is also the close relationship, observed in (\ref{eq:B1B4ij}),  between the remaining, $B^1$ and $B^4$  flux fields  in the sphere directions.   This becomes even stronger on the simple special locus, where it extends to the mixed $S^3$ and AdS$_3$ directions,  as observed in   (\ref{eq:simpIdent2}).  This extension probably does not hold for the general special locus, but, as we will discuss below, there should be some analog of this statement. 

While our work described here gives some important insights into the gravitational structure of the special locus, and  indeed suggests how one might extend  the Ans\"atze used in \cite{Ganchev:2021pgs,Ganchev:2021ewa,Ganchev:2023sth,Houppe:2024hyj} to other sphere modes in six-dimensional supergravity, our observations also raise yet more questions about the geometry of the special locus.

As we noted in the Introduction, holographic analysis \cite{Ganchev:2023sth} leads to a very simple and rather beautiful interpretation of the special locus   in terms of the  vanishing of certain vevs in the CFT, and the realization of the resulting microstrata in terms of a gas of single-particle states.   This observation nicely characterizes the algebraic constraint between the scalars, $m_{IJ}$ and $\chi^I$, embodied in (\ref{mmatrix2})  and (\ref{eigenvalue}).  At leading perturbative order, the scalars $\chi^I$ determine the $B^4$ gauge field, while the $m_{IJ}$ are shape modes of the metric (see (\ref{eq:6dmet})) and it is thus natural to see the emergence of a special shape for the $S^3$.   However, the apparent $SO(3)$ symmetry is broken by the gauge fields.     Going one step further, the Q-ball/coiffuring Ansatz determines how the other flux fields are related to the scalars $\chi^I$ and the warp factor $\Delta$, and this leads to simplifications in the components of the 2-form fields, $B$, along the sphere directions.

 From this holographic perspective, the algebraic constraints of the special locus are natural, but  the first-order system of equations, (\ref{Maxeqn1a})--(\ref{nueqn1a}), for the gauge fields is a surprise.  We have shown how this locks the fields $B^1$ and $B^2$ together, imposing a self-duality on the D1-D5 background to all orders.  It would be very interesting to see precisely what this implies for the holographic field theory at higher orders.  {\it A priori}, we were expecting a simplification related to $B^4$, and while  (\ref{eq:B1B4ij}) reflects our expectations, it does not lead to a simplification of the other components of $B^1$ and $B^2$.  Again, the surprise is the simplicity of $B^1$ and $B^2$, and not $B^4$.

There are also questions remaining as to the geometry of the special locus.   At fixed values of $(\tau, \psi, \xi)$, the metric has an $SO(3)$ isometry.  It is the gauge fields, or the KK fibration, that breaks this symmetry, and these gauge fields satisfy a first order system.  For the superstratum, some of these first-order equations were essential to creating the ambi-polar hyper-K\"ahler base geometry \cite{Ganchev:2022exf} that one associates with supersymmetry.  While we have seen that the first-order system is at the core of the self-duality of the $B^1$ and $B^2$ fields, we still do not know if they play a special role in the geometry of the metric.   Moreover, our analysis here has made no special use of the other two first-order equations, (\ref{keqn1}) and (\ref{Heqn1}), involving the three-dimensional metric functions.  These equations played an essential role in the hyper-K\"ahler properties of the superstratum base, but it remains unclear what role they play in the more general, non-supersymmetric metric structure of microstrata.  

It is, perhaps, significant  that  the two ``unused'' first-order equations involve $k$ and $H$, defined in (\ref{genmet1}) and  (\ref{FHdefns}), and these are precisely the metric coefficients of the five-dimensional spatial sections of a conformally rescaled microstratum metric  (\ref{eq:6dmet}) obtained by setting $d \tau =0$.  This suggests there may be some simplification of the five-dimensional geometry, and that $\tau$ can then be appropriately fibered over this spatial geometry. 

We leave these questions for future work, and finish by noting that we have achieved our primary goal:  determining a universal simplifying feature of the special locus that can be used to generalize to higher modes on the $S^3$.  We now know that the  D1-D5 background must be self-dual and the $S^3$ fluctuations of $B^1$ are precisely locked to those of $B^4$ when multiplied by the axion.

\vspace{1em}\noindent {\bf Acknowledgements:} 
 This work was supported in part by the ERC Grant 787320 - QBH Structure and by the DOE grant DE-SC0011687.  We would also like to thank Stefano Giusto and Rodolfo Russo for their comments on an earlier version of this paper. 
\newpage

\newpage

\begin{adjustwidth}{-1mm}{-1mm} 

\bibliographystyle{utphys}      

\bibliography{references}

\end{adjustwidth}


\end{document}